\documentclass[twocolumn]{aastex631}

\usepackage{bold-extra}

\usepackage{booktabs} 
\usepackage{longtable}
\usepackage{ltablex}
\keepXColumns % Keeps column width consistent

\usepackage{amsmath}
\begin{document}

\title{A New Age-Activity Relation For Solar Analogs That Accounts for Metallicity}

\author[0000-0002-0715-709X]{Gabriela Carvalho-Silva
}
\affiliation{Departamento de Astronomia do IAG/USP, Universidade de São Paulo, Rua do Matão 1226, 05508-090 São Paulo, SP, Brazil}
\affiliation{Astrophysics Group, University of Exeter, Exeter EX4 2QL, UK}
\email{carvalho.silva.astro@gmail.com}

\author{Jorge Mel\'endez}
\affiliation{Departamento de Astronomia do IAG/USP, Universidade de São Paulo, Rua do Matão 1226, 05508-090 São Paulo, SP, Brazil}

\author{Anne Rathsam}
\affiliation{Departamento de Astronomia do IAG/USP, Universidade de São Paulo, Rua do Matão 1226, 05508-090 São Paulo, SP, Brazil}

\author{J. Shejeelammal}
\affiliation{Departamento de Astronomia do IAG/USP, Universidade de São Paulo, Rua do Matão 1226, 05508-090 São Paulo, SP, Brazil}

\author{Giulia Martos}
\affiliation{Departamento de Astronomia do IAG/USP, Universidade de São Paulo, Rua do Matão 1226, 05508-090 São Paulo, SP, Brazil}

\author{Diego Lorenzo-Oliveira}
\affiliation{Laboratório Nacional da Astrofísica, Rua Estados Unidos 154, 37504-364 Itajubá, MG, Brazil}

\author{Lorenzo Spina}
\affiliation{INAF-Osservatorio Astronomico di Padova, Vicolo dell’Osservatorio 5, 35122, Padova, Italy}
\affiliation{INAF-Osservatorio Astrofisico di Arcetri, Largo E. Fermi 5, 50125, Firenze, Italy}

\author{D\'ebora Ribeiro Alves}
\affiliation{Departamento de Astronomia do IAG/USP, Universidade de São Paulo, Rua do Matão 1226, 05508-090 São Paulo, SP, Brazil}
\affiliation{Observatório Nacional/MCTIC, R. Gen. José Cristino, 77, 20921-400, Rio de Janeiro, RJ, Brazil}



\begin{abstract}

Determining stellar ages is challenging, particularly 
%for field stars, as classical methods rely on precise atmospheric parameters and lose precision in age
for cooler  main-sequence stars. Magnetic evolution offers an observational alternative for age estimation via the age-chromospheric activity (AC) relation. We evaluate the impact of metallicity on this relation
using near one-solar-mass stars across a wide metallicity range. We analyze a sample of 358 solar-type stars with precise spectroscopic parameters determined through a line-by-line differential technique and with ages derived using Yonsei-Yale isochrones. We measured chromospheric activity (S-index) using high-quality HARPS spectra, calibrated to the Mount Wilson system, and converted to the $R^{\prime}_{\mathrm HK}(T_{\mathrm{eff}})$ index with a temperature-based photospheric correction. Our findings show that the AC relation for $R^{\prime}_{\mathrm HK}(T_{\mathrm{eff}})$ is strongly influenced by metallicity. 
We propose a new age-activity-metallicity relation for solar-type main-sequence (MS) stars ($\log{g} \gtrsim 4.2 $) with temperatures 5370 $\lesssim$ $T_{\mathrm{eff}}$ $\lesssim$ 6530 K and metallicities from -0.7 to +0.3 dex. We show that taking metallicity into account significantly enhances chromospheric ages' reliability, reducing the residuals' root mean square (RMS) relative to isochronal ages from 2.6 Gyr to 0.92 Gyr. This reflects a considerable improvement in the errors of chromospheric ages, from 53\% to 15\%. The precision level achieved in this work is also consistent with previous age-activity calibration from our group using solar twins.

\end{abstract}

% \keywords{Observational astronomy(1145) --- Spectroscopy(1558) --- Stellar evolution (1599) --- Stellar ages(1581) --- Stellar activity(1580) --- Fundamental parameters of stars (555) --- Stellar ages (1581) --- Stellar properties(1624)}


\section{Introduction} \label{sec:intro}

Age determination is a big challenge in astronomy because of the limitations of the different methods, especially when applied to field stars. The isochronal method, for instance, can give good results for cluster stars by comparing isochrones with color-magnitude diagrams \citep[e.g.][]{2019Bossini}. However, for field stars on the main sequence, this method is quite sensitive to the precision of the stellar parameters of individual stars \citep[e.g.][]{2012Melendez}, and the achieved precision in age is usually not as good as it is for cluster stars.

Alternatively, the changes in magnetic activity along the stars' evolution can be used to estimate stellar ages. Stellar magnetic activity decreases with stellar age in low-mass stars during the main sequence (MS). Skumanich's seminal study \citep{1972Skumanich}, which relied on data from young clusters (Pleiades, Hyades), UMa moving group, and the Sun, found that activity decays with the inverse square root of the stellar age. Since then, his proposed relation has been calibrated with some of the most precise stellar ages and tested against old clusters \citep[e.g.,][]{2008Mamajek,2016Lorenzo-Oliveira,2020Gondoin}, wide binaries \citep[e.g.,][]{1991Soderblom,2008Mamajek} and field single stars \citep[e.g.,][]{2013Pace,2018Lorenzo-Oliveira}.

Little work has been developed on the effect of metallicity on chromospheric ages. \cite{1991Soderblom} argued that [Fe/H] does not impact activity, but, as pointed out in the same work, lower metallicity stars have shallower calcium lines, potentially resulting in a higher inferred activity index ($\log{R^{\prime}_{\mathrm{HK}}}$). To address this, \cite{1998Rocha-Pinto, 2005Lyra&PortodeMello, 2024SouzadosSantos} investigated these effects and proposed a metallicity correction factor. Furthermore, \cite{2016Lorenzo-Oliveira} expanded this idea by incorporating mass ($M$) and metallicity ([Fe/H]) terms into age-activity relations. Lately, those initial works have not been considered much in the literature, and in most cases, the dependence on metallicity is ignored.

In this work, we investigated the metallicity effects on the $\log{R^{\prime}_{\mathrm{HK}}}$ indices of a homogeneous sample of near one-solar-mass stars, with precisely derived stellar parameters. We discuss the observational trends with age and metallicity and propose an AC activity relation that includes a metallicity term.

\section{SAMPLE} \label{sec:data}

Our initial sample consisted of 358 stars near one-solar-mass stars -- here referred to as \textit{solar analogs} -- with stellar parameters close to solar values; as described below, it is not as strict as for solar twins. In our recent works \citep{2024Shejeelammal,2023Rathsam,2023Martos, 2018Spina}, we already derived spectroscopic parameters ($T_{\mathrm{eff}}$, $\log{g}$, $v_{\mathrm{micro}}$, [Fe/H]) for 232 of these stars. In this work, we derived the spectroscopic parameters for an additional 126 stars, presented in Table \ref{tab:maintable_params}. To ensure a homogeneous and self-consistent sample, all data were processed in a manner consistent with our previous studies.

We initially selected stars based on our primary aim of investigating the influence of metallicity on age-activity relations, those with: -0.7 $\lesssim$ [Fe/H] $\lesssim$ +0.3 dex, using the iron-to-hydrogen ratio relative to the Sun, [Fe/H], as a proxy of the overall stellar metal content; with 5100 $\lesssim$ $T_{\mathrm{eff}}$ $\lesssim$ 6600 K and $\log{g}\geq 4.2$, to select unevolved stars because, from stellar models, we see a fast radii increase at the end of the MS, impacting the surface gravity significantly. To avoid stellar multiplicity, we excluded spectroscopic binaries reported in the literature and adopted a Gaia Renormalized Unit Weight Error (RUWE) threshold of $\geq$ 1.4 to flag potential stellar multiplicity \citep{2018Lindegren}

\section{SPECTROSCOPIC DATA AND STELLAR PARAMETERS} \label{sec:data_parameters}

We used spectra from the High Accuracy Radial Velocity Planet Searcher (HARPS) instrument \citep{2003Mayor}, mounted on the ESO 3.6 m telescope, at La Silla Observatory \footnote{Based on observations collected at the European Southern Observatory under ESO programmes 0100.C-0487(A), 0100.D-0444(A), 0101.C-0379(A), 0102.C-0558(A), 0102.C-0584(A), 0103.C-0206(A), 0103.C-0432(A), 0104.C-0090(A), 0104.C-0090(B), 072.C-0488(E), 072.C-0488(E), 072.C-0513(B), 072.C-0513(D), 073.C-0784(B), 074.C-0012(A), 074.C-0012(B), 074.C-0364(A), 075.C-0202(A), 076.C-0878(A), 076.C-0878(B), 077.C-0295(D), 077.C-0364(E), 077.C-0530(A), 078.C-0833(A), 079.C-0681(A), 080.C-0712(A), 081.C-0148(A), 081.C-0148(B), 082.C-0212(A), 082.C-0212(B), 084.C-0228(A), 084.C-0229(A), 085.C-0019(A), 085.C-0063(A), 086.C-0230(A), 086.C-0284(A), 086.C-0284(A), 086.C-0448(A), 087.C-0368(A), 087.C-0831(A), 088.C-0011(A), 088.C-0323(A), 088.C-0662(A), 088.C-0662(B), 089.C-0497(A), 089.C-0732(A), 090.C-0421(A), 090.C-0849(A), 091.C-0034(A), 091.C-0853(A), 091.C-0936(A), 092.C-0579(A), 092.C-0721(A), 093.C-0062(A), 093.C-0919(A), 094.C-0797(A), 094.C-0901(A), 095.C-0040(A), 095.C-0551(A), 096.C-0053(A), 096.C-0210(A), 096.C-0460(A), 096.C-0499(A), 097.C-0021(A), 098.C-0366(A), 099.C-0458(A), 105.20AK.002, 106.215E.001, 106.215E.002, 106.215E.004, 106.21R4.001, 106.21TJ.001, 108.22CE.001, 109.2392.001, 111.24ZQ.001, 183.C-0437(A), 183.C-0972(A), 184.C-0815(A), 184.C-0815(C), 184.C-0815(E), 184.C-0815(F), 188.C-0265(A), 188.C-0265(D), 188.C-0265(F), 188.C-0265(G), 188.C-0265(H), 188.C-0265(I), 188.C-0265(J), 188.C-0265(K), 188.C-0265(L), 188.C-0265(M), 188.C-0265(N), 188.C-0265(O), 188.C-0265(P), 188.C-0265(R), 190.C-0027(A), 192.C-0852(A), 196.C-0042, 196.C-0042(D), 196.C-0042(E), 196.C-1006(A), 198.C-0836(A), 289.D-5015(A), 60.A-9036(A), 60.A-9700(G), 60.A-9709(G).}. It is a fiber-fed echelle spectrograph with a high resolving power (R=115000) and remarkable stability. 

We treated the spectral data with two approaches: To investigate the stellar activity, we considered individual spectra as data points to construct a time series for each star from our sample (see Sec. \ref{sec:chromospheric-activity}); to estimate the stellar parameters, we combined spectra from each star to enhance the final SNR ($\sim$300 to 1000) to perform a differential spectroscopic analysis (see below). 

To measure the stellar indices, we downloaded publicly available spectra from the ESO archive\footnote{\url{https://www.eso.org/sci/observing/phase3.html}} with Signal-to-Noise Ratio (SNR) higher than 20 at 550 nm. We included our recent HARPS observations of solar twins in 2023 May and June (ESO program 111.24ZQ).

\begin{figure}[ht]
    \centering
	\includegraphics[width=0.8\columnwidth]{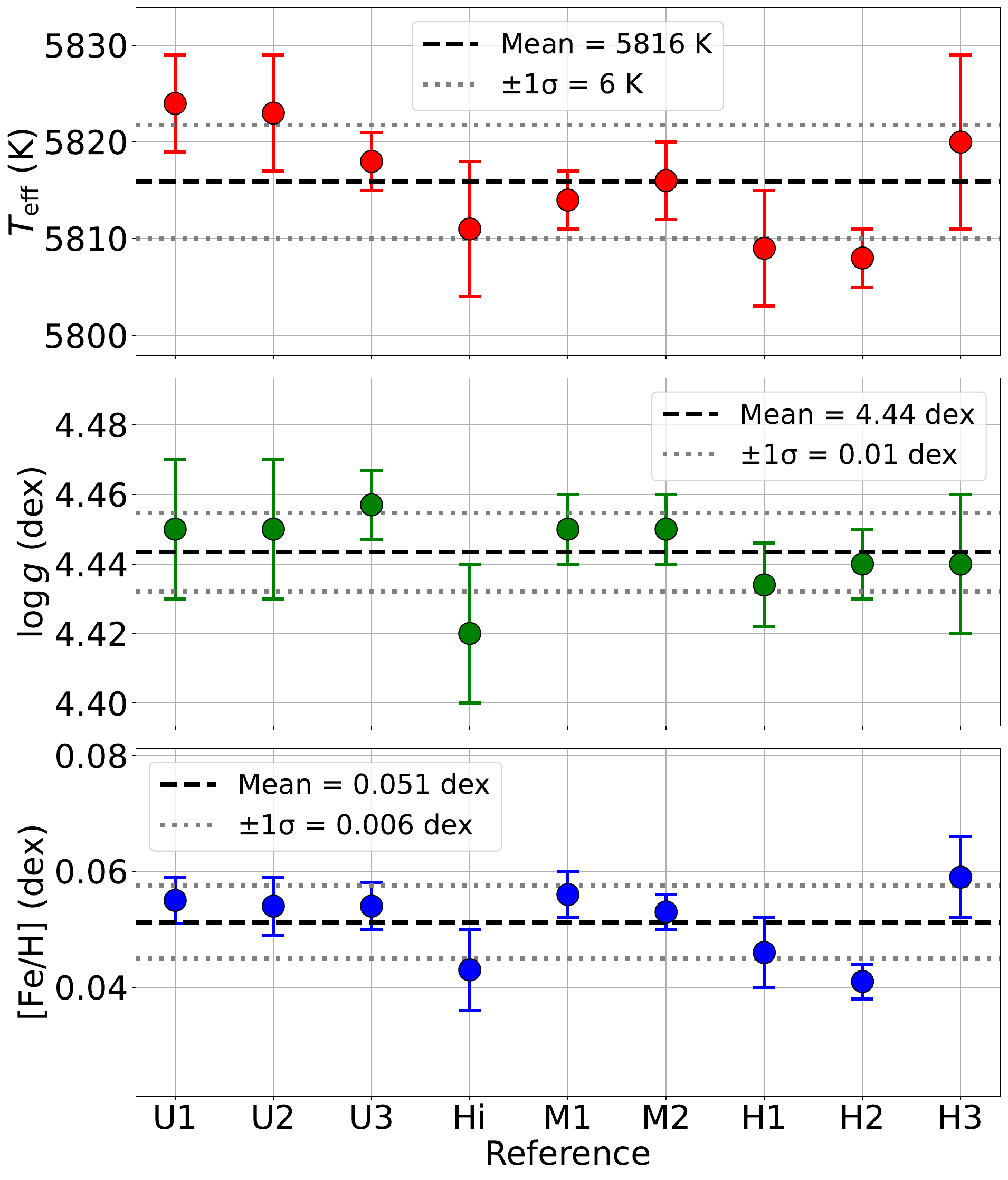}
    \caption{Differential spectroscopic parameters of 18 Sco from multiple high‐resolution instruments; UVES+VLT (U), HIRES+Keck (Hi), MIKE+Magellan (M), and HARPS (H). Horizontal lines indicate the mean value (dashed) and the $\pm 1\sigma$ dispersion (dotted). The panels show the effective temperature ($T_{\mathrm{eff}}$), the surface gravity ($\log{g}$), and the metallicity ([Fe/H]). The stellar parameters are from several literature sources based on manual line-by-line differential equivalent width analyses \citep{2013Monroe,2014Melendez,2014Ramirez,2015Nissen,2016Spina,2018Spina,2020Liu}. The last data point (Martos et al. 2025, H3) is based on a differential line-by-line relative flux analysis using machine learning.}
    \label{fig:18Sco}
\end{figure}

Regarding the consistency of the results based on different equivalent width measurements, in Figure \ref{fig:18Sco}, we plot the stellar parameters based on precise line-by-line differential analyses from different authors and spectrographs (UVES+VLT, R = 110000; HIRES+Keck, R = 67000; MIKE+Magellan, R = 65000 - 85000, HARPS+3.6m ESO, R = 115000). Despite the different spectra and different criteria used by each author, there is a good agreement among the different results. This is mainly because of two reasons: (i) the reference solar spectrum was obtained using the same spectrograph employed for the sample stars, and (ii) the measurements were performed differentially line-by-line, so that although each author may have somewhat different criteria on how to measure a particular line, the measurement is performed similarly in the star and the Sun, so that the errors are in large part canceled out. In the comparison, we also show the result of a completely automatic result (H3 points in Fig. \ref{fig:18Sco}) through a differential machine learning approach \citep[][in prep.]{2025Martos}. As can be seen, the automatic differential results are also consistent with the parameters based on the manual differential measurements.

The high precision in spectroscopic and isochronal stellar parameters comes from the combination of a homogeneous treatment of the high SNR spectra, careful manual measurements of spectral line equivalent widths, and, importantly, the \textit{differential analysis technique} \citep{2014Bedell}. The spectroscopic balance of the iron lines (Fe~\textsc{i} and Fe~\textsc{ii}) was computed to determine the atmospheric parameters using the radiative transfer code \texttt{MOOG} \citep{1973Sneden} through the automated \texttt{q}$^2$ (\texttt{qoyllur-quipu})\footnote{\url{https://github.com/astroChasqui/q2}} Python package \citep{2014Ramirez}.

The ages were estimated utilizing the Yonsei-Yale isochrones \citep{2001Yi,2002Kim}, by comparing the isochrones and the observed stellar parameters through probability distribution functions computed assuming Gaussian prior based on the input and respective errors. The inputs of the isochronal fitting are the spectroscopic parameters, $T_{\mathrm{eff}}$ and [Fe/H], GAIA DR3 \citep{2016Gaia, 2023Gaia} parallaxes and V magnitudes from the main sources at SIMBAD astronomical database \citep{2000Wenger}. We used magnesium relative to iron abundances, [Mg/Fe], as indicators of the metal enhancement due to the $\alpha$-elements \citep[see ][ for details]{2023Martos,2024Shejeelammal} to account for non-solar abundances as [Fe/H] alone does not better represent the global stellar metallicity. The resulting typical internal uncertainties in the isochronal ages and masses are $\sim$ 0.4 Gyr and 0.01 $\mathrm{M}_{\odot}$. We followed the same procedures for all sample stars and the Sun to keep the sample homogeneous to minimize systematic biases.

The isochrone grid constructed was finely spaced in metallicity (0.01 to 0.02 dex) and age (0.1 Gyr), with 279 points for each age bin. This fine sampling allowed a good resolution in mass. Overall, the grid is composed of nearly one million isochrone points. Still, for a given star, only the isochrones with metallicities within three times the error bar in [Fe/H] are used to compute the probability distribution functions in age and mass.\\Isochrone-derived ages are known to be slightly affected due to atomic diffusion effects \citep{2012Melendez,2017Dotter}. To account for this, our previous studies have shown that applying an offset of -0.04 dex to the [Fe/H] values of the Yonsei-Yale isochrones improves the age determination \citep{2014Ramirez,2018Spina}. This adjustment ensures that the solar age is recovered at 4.6 Gyr, in agreement with the well-established age of the Solar System \citep{2008Connelly,2010Amelin}. By adopting this normalization, our methodology remains consistent with the Sun and provides more reliable age estimates for solar-type stars while mitigating systematic biases in the isochrone fitting process.

\begin{table*}
\footnotesize
\centering
\begin{tabular}{rrrllllllllll}
\toprule
\text{HIP} & $T_{\text{eff}}$ [K] & $\sigma(T_{\text{eff}})$ & $\log g$  & $\sigma(\log g)$  & [Fe/H]  & $\sigma(\text{[Fe/H]})$  & $M$ [M$_{\odot}$] & $\sigma_{-}(M)$ & $\sigma_{+}(M)$  & $\text{Age}$ [Gyr] & $\sigma_{-}(\text{Age})$  & $\sigma_{+}(\text{Age})$  \\
\midrule
... & & & & & & & & &  &  &  &  \\
25616 & 5920 & 7 & 4.24 & 0.02 & -0.234 & 0.005 & 1.00 & 0.01 & 0.02 & 8.30 & 0.23 & 0.26 \\
26394 & 5992 & 4 & 4.44 & 0.01 & 0.068 & 0.003 & 1.10 & 0.01 & 0.01 & 3.70 & 0.27 & 0.25 \\
27075 & 6095 & 5 & 4.45 & 0.02 & 0.018 & 0.005 & 1.12 & 0.01 & 0.02 & 3.00 & 0.25 & 0.23 \\
27090 & 5829 & 7 & 4.24 & 0.02 & 0.273 & 0.006 & 1.11 & 0.01 & 0.02 & 6.70 & 0.26 & 0.22 \\
27244 & 6023 & 4 & 4.43 & 0.01 & 0.040 & 0.004 & 1.11 & 0.01 & 0.01 & 3.20 & 0.24 & 0.24 \\
27435 & 5733 & 4 & 4.47 & 0.01 & -0.235 & 0.004 & 0.92 & 0.01 & 0.01 & 6.30 & 0.51 & 0.26 \\
... & & & & & & & & &  &  &  &  \\
\bottomrule
\end{tabular}

\caption{Stellar parameters of 126 stars derived in this work. Parameters for the remaining 222 stars in our total sample are available in references cited in Sec. \ref{sec:data_parameters}.}\label{tab:maintable_params}
\end{table*}

\section{CHROMOSPHERIC ACTIVITY}\label{sec:chromospheric-activity}
\subsection{Time series spectral data}

The predominantly used activity proxy in the literature, the S-index \citep{1978Wilson}, is calculated based on the Ca~\textsc{ii} H (3968.470 \AA) and K (3933.664 \AA) emission lines relative to the flux in nearby continuum regions. Although the introduction of other more complex activity proxies has been suggested \citep[e.g. ][]{2024Cretignier}, the S-index is still practical and widely used by the astronomical community.

We measured the S-index and computed the corresponding $R^{\prime}_{\mathrm{HK}}$ for each selected spectrum of every star in our sample. We used the \texttt{ACTIN} \footnote{\url{https://github.com/gomesdasilva/ACTIN}} code \citep{2018GomesdaSilva,2021GomesdaSilva} to calculate the S-index, but with our calibration to the Mount Wilson System, as described in section \ref{sec:calib_S_to_SMW}.

The data filtering workflow followed these criteria: stars with fewer than 10 spectra (data points) were cut from our sample. 

We dismissed data points more than $3 \sigma$ away from the median and adopted the recalculated median of the S-index values. Temporal coverage, i.e., the timespan, is essential for obtaining a representative activity index value. As the activity cycle periods vary between stars, instead of using a fixed timespan length criterion, we visually inspected each time series to ensure that the observations were well distributed considering different periods of stellar cycles. We ended up with a sample of 324 stars.

\subsection{Calibration to the Mount Wilson System}\label{sec:calib_S_to_SMW}

We found that the S-index values calibrated to the Mount Wilson system with ACTIN, show a systematic difference with the S$_{\mathrm{MW}}$ values of the solar twins in common with \citet[hereafter LO18]{2018Lorenzo-Oliveira};  that is not justified by the band types adopted to measure the continuum and Ca~\textsc{ii} core emission; therefore, we did a new calibration to the Mount Wilson system.

We cross-matched our sample against available catalogs of S$_{\mathrm{MW}}$ in the Mount Wilson system \citep{2006Gray,1996Henry,2004Wright, 1991Duncan,2006Jenkins,2011Jenkins, 1997Piters} and found 263 objects in common with our sample. We dismissed S-index values with errors above 0.03, reducing the sample to 211 stars. 
 
 The calibration is
 \begin{equation}
    S_{\mathrm{MW}} = (1.074\pm 0.016) S_{\mathrm{Ca~\textsc{ii}}} + (0.023\pm 0.003)
\label{eq:calib_S_to_SMW}
,
\end{equation} 
where $S_{\mathrm{Ca~\textsc{ii}}}$ is the S-index measured with ACTIN and $S_{\mathrm{MW}}$ is the value calibrated to the Mount Wilson system. Its standard deviation is only $\sigma(S)$ = 0.013.

The HARPS spectrograph underwent a fiber exchange in June 2015 \citep{2015LoCurto}, MJD = 57176.5, resulting in an RV offset and changes in the continuum shape. Therefore, we investigated the impact of the HARPS upgrade on the S-index.

We found that the difference in the median S-index between the two epochs is of the same order as the standard deviation, with no clear offset. Therefore, we used data from both epochs without applying any corrections.

\subsection[transformation to logR'HK(Teff)]{ Transformation to $\log{R'_{\mathrm{HK}}}(\mathrm{T}_{\mathrm{eff}})$}
To better see the changes in the chromospheric activity indicator over time, it is convenient to transform the S-index to the logarithm of $R'_{\mathrm{HK}}$ as a function of $T_{\mathrm{eff}}$, after a $T_{\mathrm{eff}}$-based photospheric contribution. This stretches $R'_{\mathrm{HK}}(T_{\mathrm{eff}})$ in the y-axis of the age-activity diagram, because of the indicator’s higher sensitivity to small changes in chromospheric activity levels, making it possible to examine the evolutionary behavior in greater detail, as displayed in figure \ref{fig:AC_diagram}. We followed the transformation proposed in LO18 (see references therein).

We find that activity modulations in the S-index, as reflected in the standard deviation of the measurements, dominate the final error in ${R_{\mathrm{HK}}^\prime}$. These modulations contribute between 80\% and 90\% of the total uncertainty, making the S-index variability the primary driver of the final error.

All activity proxies and related parameters are listed in Table \ref{tab:activity_proxies}

\begin{table*}
\centering
\begin{tabular}{rllllrlllll}
\toprule
\text{HIP} & $\langle S_{\rm{Ca~\sc{ii}}} \rangle$ & $\sigma(S_{\text{CaII}})$ & $\langle S_{\text{MW}} \rangle$ & $\sigma(S_{\text{MW}})$ & $N_{\text{obs}}$ & $t_{\text{span}}$ [d] & $\langle \log R'_{\text{HK}} \rangle$ & $\sigma(\log R'_{\text{HK}})$ & $\langle \log R'_{\text{HK, B-V}} \rangle$ & $\sigma(\log R'_{\text{HK, B-V}})$ \\
\midrule
... & & & & & & & & &  &   \\
25616 & 0.1256 & 0.0020 & 0.1579 & 0.0022 & 40 & 7316 & -5.039 & 0.016 & -5.136 & 0.019 \\
25670 & 0.1367 & 0.0048 & 0.1699 & 0.0051 & 62 & 2865 & -5.025 & 0.031 & -5.038 & 0.030 \\
26394 & 0.1273 & 0.0020 & 0.1597 & 0.0022 & 80 & 2278 & -4.998 & 0.015 & -4.941 & 0.014 \\
27075 & 0.1279 & 0.0020 & 0.1603 & 0.0021 & 6 & 2348 & -4.958 & 0.014 & -4.958 & 0.014 \\
27090 & 0.1246 & 0.2913 & 0.1569 & 0.3128 & 8 & 2974 & -5.085 & 0.421 & -4.873 & 0.370 \\
27244 & 0.1288 & 0.0016 & 0.1614 & 0.0018 & 40 & 2354 & -4.975 & 0.011 & -4.937 & 0.011 \\
27435 & 0.1401 & 0.0015 & 0.1735 & 0.0016 & 108 & 2514 & -5.014 & 0.009 & -4.826 & 0.007 \\
... & & & & & & & & &  &  \\
\bottomrule
\end{tabular}

\caption{Activity proxies and related parameters.}\label{tab:activity_proxies}
\tablecomments{Tables 1 and 2 are published in their entirety in the machine-readable format. A portion is shown here for guidance regarding their form and content.}
\end{table*}

\section{RESULTS AND DISCUSSION}\label{sec:results}

\begin{figure*}[ht]
    \centering
	\includegraphics[width=0.9\textwidth]{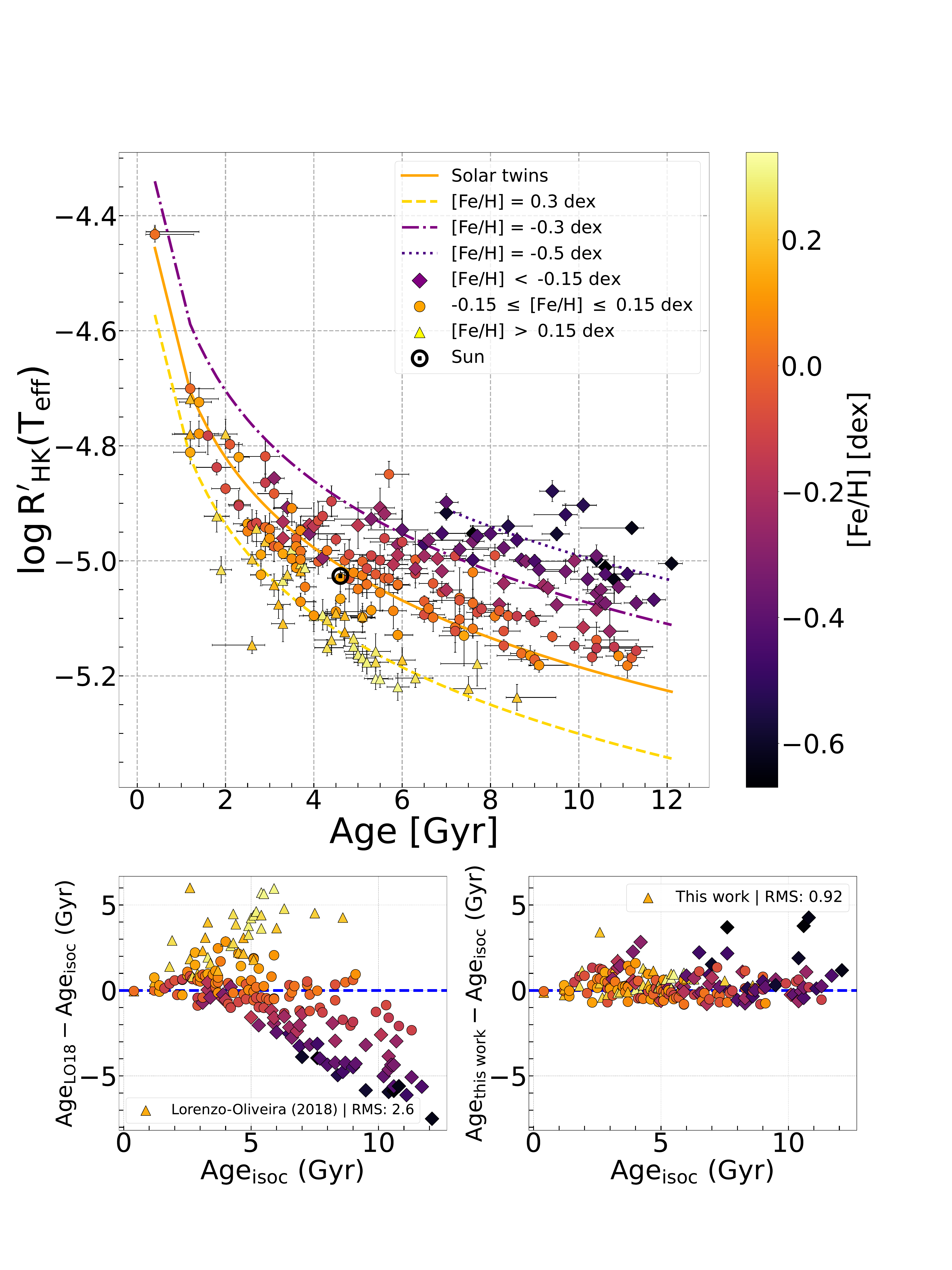}
    \caption{Age-chromospheric activity diagram: activity proxy $R^{\prime}_{\mathrm{HK}}$ against age color-mapped by metallicity as [Fe/H]. Triangles are metal-rich stars, diamonds are metal-poor stars, and circles are stars with solar metallicity. The Sun is shown with its usual symbol in black. The lines represent our proposed age-activity-metallicity relation (equation \ref{eq:Age_ACrelation_teff}) for different values of [Fe/H]. The bottom panels are the residuals between the isochronal ages and those determined by the LO18 relation based on solar twins (left panel) and those determined by our metallicity-dependent relation (right panel).}
    \label{fig:AC_diagram}
\end{figure*}

The age-activity relation for solar twins in LO18 does not hold for metallicities that depart from solar, as shown in the residuals in the bottom left panel in Fig. \ref{fig:AC_diagram}. Previous works have pointed out this dependence and proposed either a metallicity correction \citep[][hereafter RP98]{1998Rocha-Pinto} or a multiparametric relation \citep[][LO16]{2016Lorenzo-Oliveira}. Without the metallicity correction, the chromospheric ages are underestimated for metal-poor stars and overestimated for metal-rich stars.

We constructed an age-chromospheric activity diagram with $\log{R}^\prime_{\mathrm{HK}}$ values and isochronal ages. In Fig. \ref{fig:AC_diagram}, we can see the strong dependency of the activity indicator with stellar metallicity. The color mapping in [Fe/H] is crucial for discerning the stellar magnetic evolution of stars with different properties. The Sun is not used for calibration, but it is presented for visual comparison with other stars.

We applied cuts to our sample to focus on main-sequence stars at intermediate to old ages. In addition to the restriction on surface gravity to $\log{g} \geq 4.2$, we removed stars with a standard deviation in the S-index greater than $0.01$ -- using a $5\sigma$-clipping method based on the mean. A high standard deviation in the S-index indicates high amplitude and/or complex cycles, which can reflect poorly constrained mean activity levels. As a result, old, highly active stars were treated as outliers. Furthermore, since we fixed the age term (as explained below), the reduction in the number of young stars did not negatively impact our results. After applying all cuts, we obtained a final sample of 234 stars.

We performed different fittings using several functions. When considering only age and a constant term, both parameters gave significant values ($>10\sigma$). When a metallicity term was added, both age and metallicity were highly significant, with the age term showing increased significance. The assumption that stars follow the same decay of $R^{\prime}_{\mathrm{HK}}$ independent of metallicity, is inconsistent with the observed data and with our unprecedented precise measurements and careful sample selection.

The age term was highly significant in all our fits and consistent with the solar twins' values. To evaluate the metallicity impact, we assume that the decay relation in time is the same as for solar twins in the case of solar-metallicity stars. Therefore, we fixed the age parameter at $-1.92$ as in LO18. This improved the fitting performance as it no longer had to deal with balancing the high scattering beyond 2 Gyr and fewer points in younger ages. We could estimate the dependence in metallicity with 37-$\sigma$ confidence, showing a clear strong metallicity dependence.

We added mass and effective temperature terms to the function while keeping the metallicity term. Both showed similar significance levels $\approx 4\sigma$. When adding both terms, $T_{\mathrm{eff}}$ and $M$, the rms changed from 0.93 to 0.84 Gyr. These additions make the relation more complex without significantly improving the fitting, as it did not present big reductions in the rms of the residuals. Furthermore, both effects are entangled due to degeneracies between terms. Thus, we opted for a simpler form accounting for metallicity, as the improvement due to the addition of multiple terms was small. With the metallicity term alone, the rms of chromospheric ages improved from $\sim 2.6$ Gyr to $\sim 0.92$ Gyr. Our proposed relation is presented in a simple and practical form, based on accessible measurements:
\begin{multline}
    \log{Age} = 
    -\:1.92\: \log{R^{\prime}_{\mathrm{HK}}(T_{\mathrm{eff}})} \\
    - 0.74 (\pm 0.02)\: \mathrm{[Fe/H]} + 0.046 (\pm 0.005)
    \label{eq:Age_ACrelation_teff}
.
\end{multline}

To quantify the impact of the improvement in errors of chromospheric ages due to the inclusion of the metallicity term, we analyzed those stars with metallicities different from solar; $\mathrm{[Fe/H]} \leq -0.15$ dex and $\mathrm{[Fe/H]} \geq 0.15$ dex. The mean difference between chromospheric and isochronal ages is 3 Gyr compared to 0.8 Gyr when accounting for metallicity corrections. This corresponds to an improvement in the mean relative errors from 53\% to 15\%.

\begin{figure}[ht]
    \centering
	\includegraphics[width=0.8\columnwidth]{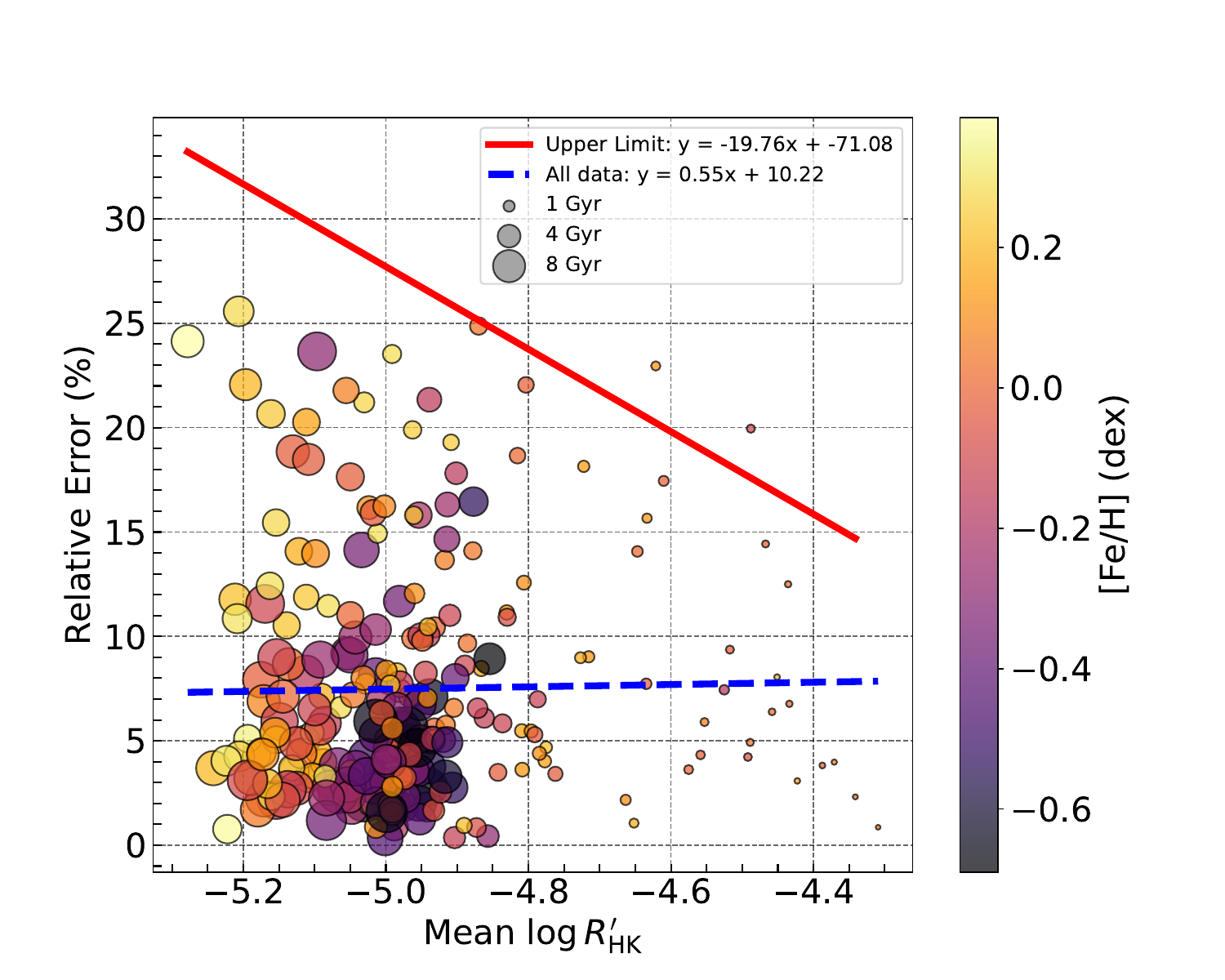}
    \caption{Relative percentage error in chromospheric ages based on individual measurements versus the mean chromospheric activity index, $\log R^{\prime}_{\rm HK}$. Each circle represents a star, where the values are computed from three randomly selected observations per star. Sizes represent their mean age (in Gyr), and colors indicate metallicity ([Fe/H]). The solid red line shows the linear regression fitted to the upper envelope (highest relative error per bin), while the blue dashed line is a linear regression across all data points.}
    \label{fig:age_dispersion}
\end{figure}

The uncertainty of derived chromospheric ages based on only a few H and K measurements highly depends on stellar variability since these snapshots may not accurately represent the mean activity level. Therefore, precise and accurate measurements, obtained with a long time baseline, are essential. To estimate the possible error due to the use of only one or just a few measurements, in Fig. \ref{fig:age_dispersion} the relative percentage error in chromospheric ages based on three randomly selected individual measurements versus the mean chromospheric activity index, $\log R^{\prime}_{\rm HK}$. We present the upper envelope (red solid line) and the mean trend (blue dashed line). For cases with only a couple of H and K data points, we recommend including in the quadratic sum of the error the highest expected dispersion defined by the upper envelope. In contrast, if more than two measurements are available, spaced over timescales of several months to years, we advise using the quadratic sum of the expected dispersion considering all data, which implies a relative error of $\sim 8 \%$.

The metallicity factor presented in our relation agrees qualitatively with other works (RP98, LO16), but our relation has a higher significance. In both works, they used the $\log{R^{\prime}_{\mathrm{HK}}}(B-V)$, an activity indicator similar to the one used in this work, but with photospheric correction based on the B$-$V color index. Moreover, past works had to deal with low precision in the stellar parameters of single stars, the sparsity of the ages of open clusters, and open cluster data limited up to 6 Gyr. In our work, we have for the first time an activity-age diagram with high-precision stellar parameters that allow us to see the metallicity effect in detail. The usage of a photospheric correction based on effective temperature was important to break part of the degeneracy that is present in the classical photospheric correction based on color-index (see a detailed discussion in LO18). 

We acknowledge that the color-index-based correction is broadly used, and for this reason, we additionally fitted an age-activity relation for $\log{R^{\prime}_{\mathrm{HK}}}(B-V)$. In that case, it is better to add a temperature term. We compared this AC relation with those of RP98 and LO16. As can be seen in Fig. \ref{fig:comp_AC_bv}, the metallicity effect remains. As both works used $\log{R^{\prime}_{\mathrm{HK}}}(B-V)$, to perform the comparison, we calculated the $\log{R^{\prime}_{\mathrm{HK}}}(B-V)$ for all stars in our final sample.

\begin{figure}[ht]
    \centering
	\includegraphics[width=1.\columnwidth]{comp_bv_plot.pdf}
    \caption{Comparison between isochronal ages and the AC relations from \cite{1998Rocha-Pinto,2008Mamajek,2016Lorenzo-Oliveira} and this work. Color-mapping is the same as Fig. \ref{fig:AC_diagram}}
    \label{fig:comp_AC_bv}
\end{figure}

 Our alternative new relation for age derivation from the logarithm of $R^{\prime}_{\mathrm{HK}}$ with color-based photospheric correction is:

\begin{multline}
    \log{Age} = 
    -\:1.92\: \log{R^{\prime}_{\mathrm{HK}}(B-V)} \\
    -0.79 \left (\pm 0.02\right )\: \mathrm{[Fe/H]} - 6\times 10^{-4} \left ( \pm 2\times 10^{-5}\right) T_{\mathrm{eff}}\\
    + 3.7 (\pm 0.1)
    \label{eq:Age_ACrelation_bv}
.
\end{multline}

\section{CONCLUSIONS}
In this work, we analyze 358 about one-solar-mass main-sequence stars in the broad metallicity range of -0.7 $\lesssim$ [Fe/H] $\lesssim$ +0.3 dex and with precise stellar parameters. We used HARPS spectra and measured activity indices ($S_{\mathrm{Ca~II}}$). We compared data from before and after the HARPS 2015 fiber upgrade and verified no significant variation between these measurements. We calibrated the measured S-index to the Mount Wilson system with our up-to-date calibration, and the $\log{R'_{\mathrm{HK}}}(T_{\mathrm{eff}})$ indices were obtained.

We showed that the AC relation highly depends on metallicity. According to our relation, the predicted errors due to different metallicities for typical about one-solar-mass main-sequence stars around solar age with [Fe/H] = +0.3 dex, relative to solar metallicity, amount to a few Gyrs; the impact increases for older ages and metallicities departing from solar values. Our results agree qualitatively with \cite{1998Rocha-Pinto} and \cite{2016Lorenzo-Oliveira}, but our work significantly improves the robustness of age-activity relations.

This conclusion aligns with previous independent studies that have analyzed chromospheric activity on large stellar samples and other chromospheric indicators, particularly those utilizing data from surveys such as LAMOST \citep[][and references therein]{2024Zhang}.  

Our proposed relation can impact future works in different areas, such as determining the ages of exoplanet host stars or studying the formation history of the Milky Way through more accurate stellar ages.

\begin{acknowledgements}

G.C.S. thanks the FAPESP PhD fellowships 2021/01303-3 and 2023/16319-8. We thank the support of FAPESP (2018/04055-8, 2022/05833-0, 2022/10325-3). This work has made use of data from the European Space Agency (ESA) mission {\it Gaia} (\url{https://www.cosmos.esa.int/gaia}), processed by the {\it Gaia} Data Processing and Analysis Consortium (DPAC, \url{https://www.cosmos.esa.int/web/gaia/dpac/consortium}). Funding for the DPAC has been provided by national institutions, in particular, the institutions participating in the {\it Gaia} Multilateral Agreement. This research has made use of the SIMBAD database, operated at CDS, Strasbourg, France.
\end{acknowledgements}

\vspace{5mm}
\facilities{ESO:3.6m, Gaia.}
\software{ACTIN \citep{2018GomesdaSilva, 2021GomesdaSilva}, Astropy \citep{astropy:2013, astropy:2018, astropy:2022}, Astroquery \citep{astroquery}, IRAF \citep{iraf:1, iraf:2}, Matplotlib \citep{matplotlib}, MOOG \citep{1973Sneden}, NumPy \citep{numpy}, pandas \citep{pandas},  q2 \citep{2014Ramirez}.}


% \software{astropy \citep{2013A&A...558A..33A,2018AJ....156..123A}
%           }



\bibliography{sample631}{}
\bibliographystyle{aasjournal}

% \allauthors


\end{document}